# Three-dimensional topological valley photonics

Wenhao Li[1,2,3,4], Qiaolu Chen[1,2,3,4], Ning Han[1,2,3,4], Xinrui Li[1,2,3,4], Fujia Chen[1,2,3,4], Junyao Wu[1,2,3,4], Yuang Pan[1,2,3,4], Yudong Ren[1,2,3,4], Hongsheng Chen[1,2,3,4],[*] Haoran Xue[5],[†] and Yihao Yang[1,2,3,4][‡]

[1] *Interdisciplinary Center for Quantum Information State Key Laboratory of Extreme Photonics and Instrumentation ZJU-Hangzhou Global Scientific and Technological Innovation Center, Zhejiang University, Hangzhou, China*

[2] *International Joint Innovation Center The Electromagnetics Academy at Zhejiang University. Zhejiang University, Haining, China*

[3] *Key Lab. of Advanced Micro/Nano Electronic Devices Smart Systems of Zhejiang Jinhua Institute of Zhejiang University, Zhejiang University, Jinhua, China*

[4] *Shaoxing Institute of Zhejiang University, Zhejiang University, Shaoxing, China and*

[5] *Department of Physics, The Chinese University of Hong Kong, Shatin, Hong Kong SAR, China*

**Abstract**

Topological valley photonics, which exploits valley degree of freedom to manipulate electromagnetic waves, offers a practical and effective pathway for various classical and quantum photonic applications across the entire spectrum. Current valley photonics, however, has been limited to two dimensions, which typically suffer from out-of-plane losses and can only manipulate the flow of light in planar geometries. Here, we have theoretically and experimentally developed a framework of three-dimensional (3D) topological valley photonics with a complete photonic bandgap and vectorial valley contrasting physics. Unlike the two-dimensional counterparts with a pair of valleys characterized by scalar valley Chern numbers, the 3D valley systems exhibit triple pairs of valleys characterized by valley Chern vectors, enabling the creation of vectorial bulk valley vortices and canalized chiral valley surface states. Notably, the valley Chern vectors and the circulating propagation direction of the valley surface states are intrinsically governed by the right-hand-thumb rule. Our findings reveal the vectorial nature of the 3D valley states and highlight their potential applications in 3D waveguiding, directional radiation, and imaging.

## INTRODUCTION

Inspired by the success of using valleys to manipulate electrons in two-dimensional (2D) materials, significant progress has been made in exploiting the valley degree of freedom in photonics, supplementing conventional ones such as amplitude, phase, polarization, and frequency, to control the flow of light [1–4]. Traditionally, a 2D valley photonic crystal (VPC) contains a single pair of valleys located at the corners of the 2D Brillouin zone (BZ). The Bloch modes around these valleys are described by two-band massive Dirac models, related by time-reversal symmetry and exhibiting opposite scalar band geometric properties, such as Berry curvature ($\Omega$). This opposition leads to various physical behaviors that differ between the two valleys, known as valley-contrasting physics [1]. In photonics, two contrasting phenomena related to valley-contrasting physics have been extensively studied. One is the formation of vortex states with valley-locked chirality, generated by the valley-contrasting orbital magnetic moment. The other phenomenon involves valley edge states, which are protected by the valley Chern number—a half-integer scalar topological invariant obtained by integrating the valley-contrasting Berry curvatures [5–19].

Valley photonics presents a readily accessible and effective method for achieving topologically robust photonic devices suitable for real-world applications [5,7,13,20]. Unlike other topological phases, such as photonic Chern insulators and photonic analogues of topological insulators, VPCs are more experimentally feasible because they do not necessitate breaking time-reversal symmetry or incorporating synthetic spin-orbital coupling. Instead, achieving VPCs simply requires a reduction of inversion symmetry, which guarantees the presence of local nonzero Berry curvature. These advantages make VPCs versatile across broad frequency ranges and adaptable to various material platforms, suitable for both classical and quantum photonics, enabling various applications, including lasers [10,11], on-chip communication [9,14], and quantum optics [8].

While valley photonics has attracted tremendous attention, all previous studies were limited to 2D systems [2,3], where the photonic confinement along the third dimension is usually obtained in a nontopological way, such as the total internal reflection. These 2D valley systems typically support valley states with a single polarization, suffer from out-of-plane losses, and can only manipulate the flow of light in 2D planar geometries. Valley photonics with full confinement of light and full control of the flow of light can only be achieved by the complete photonic bandgap (in which light with any polarization is forbidden to propagate in any direction in three dimensions), which, however, remains a significant challenge.

Here, we introduce a theoretical and experimental framework of three-dimensional (3D) topological valley photonics with a complete photonic bandgap. Our 3D framework goes beyond the binary valley degree of freedom and scalar valley-contrasting physics pictures in the 2D case. Specifically, instead of a single pair, our 3D model involves triple pairs of valleys at the corners of the 3D BZ. These pairs are differentiated by band geometric quantities along distinct directions, fully leveraging the vectorial nature of these quantities in three dimensions and introducing a vectorial valley-contrasting physics perspective. Using the pump-probe measurement, we verify the vectorial valley-contrasting Berry curvature by selectively exciting and observing the chiral Bloch modes through the vectorial chiral source array. Also, we demonstrate the robust canalized valley surface states on all surfaces of a specifically shaped sample—a capability not previously achieved by any 3D photonic topological insulators, and reveal the right-hand-thumb rule relating the direction of the valley Chern vectors and the propagation loops of valley surface states.

## RESULTS

We start from a 3D photonic crystal with lattice constant $a = 20$ mm, as illustrated in Fig. 1(a), and the lattice vectors are denoted as $a_1 = (a/2, a/2, 0)$, $a_2 = (0, a/2, a/2)$ and $a_3 = (a/2, 0, a/2)$. The photonic crystal is a sort of photonic "diamond" with both nearest-neighbor and next-nearest neighbour couplings, in which the carbon atoms and the bonds are replaced by perfect electric conductor (PEC) spheres and rods, respectively. Each face-centered-cubic (fcc) cell contains two spheres with diameters D1 and D2 (Fig. 1(a), right panel), respectively. Each nearest sphere is connected by PEC rods with diameter $B_1 = 2.4$ mm. Additionally, we connect the next-nearest spheres by PEC rods with diameter $B_2 = 1.2$ mm to suppress high-order modes, so that we can obtain a clean bandstructure where the topological bands are separated from the trivial bands [21].

The band structure of the 3D photonic crystal is shown in Fig. 1(b). When $\Delta D = 0$ ($\Delta D \equiv |D_1 - D_2|$), the inversion symmetry is preserved, and the photonic crystal hosts nodal lines along $W_\alpha^\pm$-X ($\alpha = x, y, z$), forming intersecting nodal chains. The momentum of the six valleys are $(\pm\pi/a, 2\pi/a, 0)$, $(0, \pm\pi/a, 2\pi/a)$, and $(2\pi/a, 0, \pm\pi/a)$ for $W_x^\pm$, $W_y^\pm$, and $W_z^\pm$, respectively (Fig. 1(c)). When $\Delta D \neq 0$, however, the inversion symmetry is broken, and the nodal line degeneracies are gapped out to form a complete photonic bandgap. The 3D gapped phase has six valleys at $W_\alpha^\pm$, which can be grouped pairwise, due to the time-reversal symmetry. We plot the distributions of $H_z$'s phase and Poynting vector in the x-y plane for the eigenstates at $W_z^\pm$ of the lower band in Fig. 1(d). One can see that both eigenstates display typical vortex profiles but opposite chirality, as required by the time-reversal symmetry. The Berry curvature at $W_z^\pm$ with nonzero component only in the $k_x$-$k_y$ plane verifies the vectorial valley-contrasting physics (Fig. 1(e)) [21].

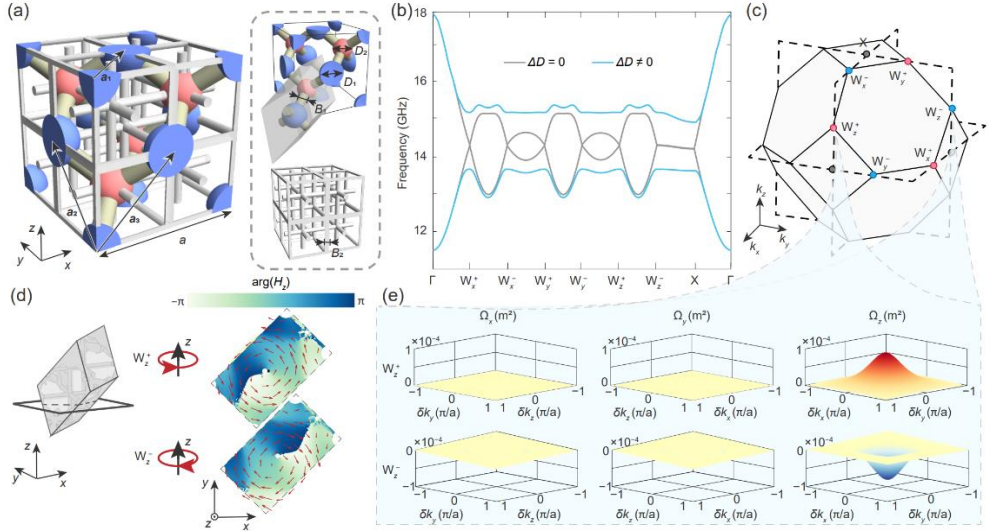

FIG. 1. 3D VPC characterized by vectorial Berry curvatures. (a) Cubic cell. The left panel indicates the cubic cell consisting of two sublattices (denoted by blue and red spheres) with lattice constant a, and lattice vectors $a_1$, $a_2$, and $a_3$. The cubic cell consists of a diamond-like lattice and a metallic mesh that connects two next-nearest-neighbour spheres in each sublattice (right panel). Gray fcc unit cell consists of two spheres. A cubic cell consists of four fcc unit cells. (b) Band structure of the 3D VPC. The gray and blue lines indicate the band structures when $\Delta D = 0$ ($D_1 = D_2 = 5$ mm) and $\Delta D \neq 0$ ($D_1 = 6.4$ mm, $D_2 = 3.6$ mm), respectively. (c) 3D BZ of the fcc unit cell. The six valleys are denoted by red and blue dots. The nodal lines are denoted by the dashed lines. (d) Simulated eigenstates on the (001) surface for the first band at $W_z^\pm$ valleys. The left panel shows the position in the $z$ direction within a fcc unit cell. Red arrows indicate the Poynting vector, and the colors indicate the phase of $H_z$. (e) Simulated Berry curvature components of the first band at $W_z^\pm$.

By applying the $k \cdot p$ method, we obtain the effective Hamiltonian near each valley:

$$\mathbf{H}_{W_\alpha^\pm}(\mathbf{q}) = \pm v_D \sigma_x q_\beta + v_D \sigma_y q_\gamma + m \sigma_z, \quad (1)$$

where $(\alpha, \beta, \gamma) = (x, y, z)$, $(y, z, x)$, or $(z, x, y)$, $\mathbf{q} = (q_x, q_y, q_z)$ represents the momentum offset from each valley, $v_D$ is the group velocity, $\sigma_{x,y,z}$ are Pauli matrices, and $m = (\omega_1 - \omega_2)/2$ is the mass term induced by breaking the inversion symmetry, with $\omega_{1,2}$ the angular frequencies of the two bands at the valleys [21].

The different chirality of Bloch modes at each pair of valleys indicates the valley-chirality locking property. Consequently, the state at each valley can be selectively excited using a chirality-matched source array. To implement the phase winding on $x$-$y$, $y$-$z$, and $x$-$z$ planes in experiments, the vectorial chiral source array is configured with four sources with the relative positions at $(0, 0, 0)$, $(-a/2, -a/2, 0)$, $(-a/2, 0, -a/2)$, and $(0, -a/2, -a/2)$. Taking valley states at $W_y^\pm$ as an example, the relative phase delay of each source is set to $(0, \pi/2, \pi, 3\pi/2)$ for $W_y^-$ and $(0, -\pi/2, -\pi, -3\pi/2)$ for $W_y^+$, to form a clockwise or counterclockwise phase winding in the $x$-$z$ plane (see Fig. 2(a)). The excited valley states will outcouple to the free space radiation and, thus, the momentum of the excited valley state can be judged from the outcoupling direction (see Figs. 2(b) and (c)). As shown in Fig. 2(d), the measured outcoupling fields at 13.8 GHz exhibit different directions: for the clockwise (counterclockwise) phase winding, the direction is −32.9° (+32.9°), corresponding to $W_y^-$ ($W_y^+$), while for the case without phase winding, the direction is 0°, corresponding to X, since the states at X coexist with valley states at the same frequency [22]. The above phenomena can be well understood from the momentum-space analysis, where the valley state and the free-space radiation should share the same tangential momentum parallel to the boundary (e.g., the (001) surface) (see Fig. 2(b)). Furthermore, the directionality of the Bloch modes can be verified by the far-field radiation patterns, where the outcoupling directions agree well with the calculated results (see Fig. 2(e)). Note that, unlike the 2D case where the radiation pattern is restrained only in the 2D plane and diverges in the vertical direction [7], the radiated beams in Fig. 2(e) are single-directional in 3D space, which provides the opportunity for the directional antenna design.

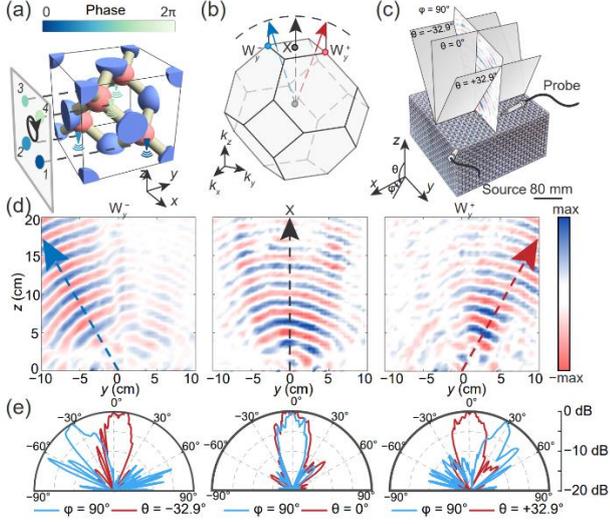

FIG. 2. Selective excitation of vectorial valley states. (a) Illustration of the vectorial chiral source array. The relative positions of four sources are denoted by four beacons. The colors of the beacons indicate the relative phase delay for $W_y^-$. The arrow denotes the phase winding of $W_y^-$. The next-nearest-neighbor coupling rods are removed for clarity. (b) Momentum-space analysis. The blue, black, and red dots indicate $W_y^-$, X, and $W_y^+$, respectively. The arrows represent the outcoupling directions. The dashed black arc denotes the free-space light dispersion at 13.8 GHz. (c) Experimental setup. Four scanned planes are labeled as $\varphi = 90°$, $\theta = -32.9°$, $\theta = 0°$, and $\theta = +32.9°$. (d) Measured electric field distributions of outcoupling waves in the y-z plane. The dashed blue, black, and red arrows indicate the theoretical refraction direction, which is $-32.9°$, $0°$, and $+32.9°$ with respect to the z-axis, respectively. (e) Measured outcoupling far-field radiation patterns at $W_y^-$, X, and $W_y^+$ in planes in (c), respectively.

The band topology of our 3D VPC can be manifested by the local Berry curvature around each valley [9], where our 3D VPC possesses vectorial Berry curvatures, i.e., $\mathbf{\Omega} = (\Omega_x, \Omega_y, \Omega_z)$ (see Fig. 1(e)). By integrating the Berry curvature around each valley,

$$\mathbf{C}_v = \frac{1}{2\pi} \int_S \mathbf{\Omega}(\mathbf{q}) d\mathbf{S}, \quad (2)$$

where $S$ is a planar region around valleys normal to $q_x$, $q_y$, and $q_z$, respectively. We calculate the valley Chern vector (VCV) [$\mathbf{C}_v = (C_{vx}, C_{vy}, C_{vz})$], which is ($\pm 1/2$, 0, 0), (0, $\pm 1/2$, 0), and (0, 0, $\pm 1/2$) for $W_x^\pm$, $W_y^\pm$, and $W_z^\pm$ valleys, respectively [21].

For topological valley photonics, there are generally two approaches to the topological boundary states. One approach is forming a domain wall between two valley crystals with opposite valley Chern numbers, with the valley kink states appearing at the domain wall. The other approach is modifying the on-site boundary potentials of a valley crystal, with the chiral boundary states existing at the external surface, enabling compact photonic devices [1,23].

Here, we adopt the latter approach to facilitate the sample fabrication and experimental characterization. We first derive the general condition that the topological chiral surface states can exist on our 3D VPC. Without loss of generality, we consider the (001) surface [24], and the topological surface states, denoted as $\Phi$, satisfying the equation $\mathbf{H}\Phi = \omega\Phi$, as well as the general boundary condition $\hat{\mathbf{M}}\Phi = \Phi$. Here, $\hat{\mathbf{M}} = \tau_z (\sigma_y \sin\theta_V + \sigma_z \cos\theta_V)$, where $\theta_V$ phenomenologically represents the influences of the boundary condition on the surface states [13,20]. One can find that only when $\theta_V = \pi/2$, $\Phi$ is a chiral surface state determined by the VCV with a linear dispersion $\omega = 2v_D \mathbf{S} \cdot (\mathbf{q} \times \mathbf{C}_v)$ [$\mathbf{S}$ is the vectorial Miller index of surface, e.g., ($-1$, 0, 0) for the ($\bar{1}$00) surface] [21]. To achieve the desired boundary condition in practice, we cover the (001) surface with a metallic plate and tune the diameter of the outmost metallic sphere, $D_3$, and the distance between the plate and the outmost metallic sphere, $h$ (see Figs. 3(a) and (b)). Note that the metallic plate features perforations for inserting the source and probe, with a hole interval of $a/2$. We find that when $h = 6.25$ mm and $D_3 = 4$ mm, the boundary condition can be satisfied [21]. Consequently, a linearly-dispersed valley-locked surface state traverses the bandgap (see Figs. 3(c) and (d)).

We then perform experiments to demonstrate the chiral surface states from the VCV. The measured surface transmission, depicted in Fig. 3(e), maintains high within the complete photonic bandgap, while the bulk transmission is substantially lower, which verifies the existence of the surface states. We also map out the surface field distribution (Fig. 3(f)) and obtain its Fourier-transformed spectrum along $L_4$-$W_x^-$-$L_1$ (Fig. 3(g)). We can indeed observe a nearly-linear surface dispersion in the bandgap. Note that the chiral surface states exhibit canalization behaviour; that is, the surface states propagate diffractionlessly along the +y direction, corresponding to the $W_x^-$ valley. This canalization phenomenon is due to the flat isofrequency contour of the surface states, which has been confirmed both experimentally and numerically (Fig. 3(h)) [25–27]. The canalization has previously been explored at the topological transition point between open hyperbolic and closed elliptical isofrequency contours, as in the twisted van der Waals bilayers [26]. From the equation of the surface states aforementioned, one can see that the surface dispersion is independent of the wavevector vertical to the propagation direction (denoted as $k_x$ ($k_y$)), indicating that the group velocity of surface states preserves for different $k_y$ ($k_x$) around each valley, which leads to the flat isofrequency contours. Thus, the canalization is an intrinsic property of our surface states, presenting promising opportunities for imaging and lensing [21,26,28].

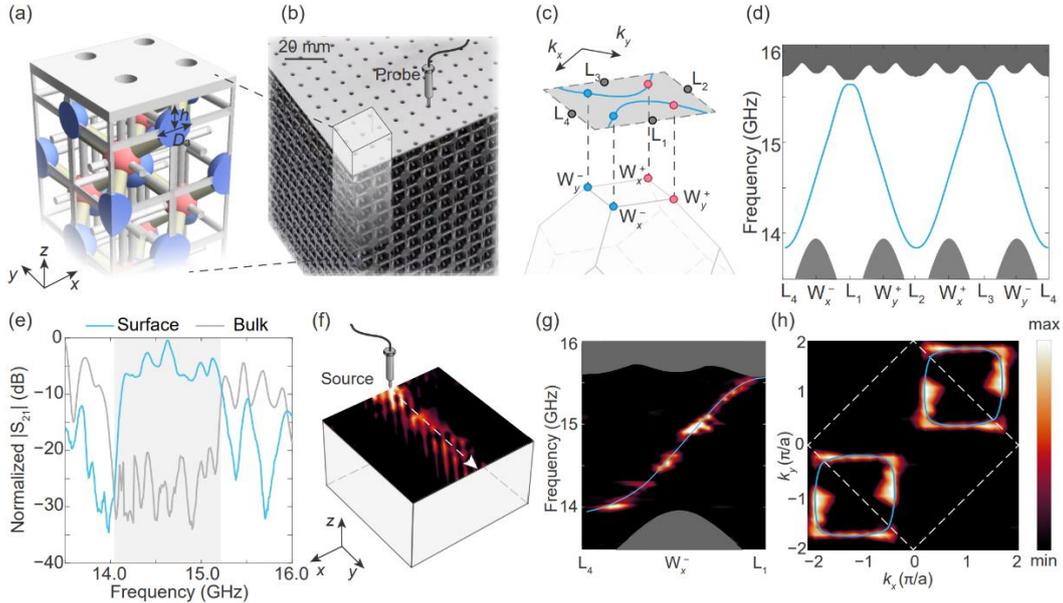

FIG. 3. Observation of the canalized chiral valley surface states. (a) 3D VPC with the modified (001) surface. The surface is covered by a metallic plate. $h$ and $D_3$ are key parameters to modify the boundary condition. (b) Experimental setup. (c) Projective surface BZ onto the (001) surface. The blue lines indicate the simulated isofrequency contour of the surface dispersion at 14.6 GHz. The gray region represents the projective BZ. (d) Simulated surface dispersion. The gray areas and the blue line represent the bulk states and the surface states, respectively. (e) Normalized measured transmission. The shadowed area denotes the bulk bandgap. (f) Measured field distribution at 14.6 GHz. The dashed line represents the propagation direction of the canalized surface states. (g) Measured (colormap) and simulated (blue line) surface dispersion around $W_x^-$ valley. (h) Measured (colormap) and simulated (blue lines) isofrequency contour of surface dispersion at 14.6 GHz. The dashed square represents the first BZ.

Unlike the existing 3D topological photonic insulators [29–34] in which the topological surface states can only exist on certain surfaces, the 3D VPCs with certain shapes can support the chiral valley surface states on all sample surfaces. Obviously, our fabricated sample can support the chiral valley surface states on all (100), (010), and (001) surfaces, with the propagation direction locked to each valley. For instance, the surface states at $W_x^-$ valley could propagate along +y direction on the (001) surface, and subsequently along −z direction on the (010) surface, without experiencing intervalley scattering, then along −y direction on the

($00\bar{1}$) surface and $+z$ direction on the ($0\bar{1}0$) surface, which forms a closed loop. One can see that the direction of VCVs and the circulating direction of the valley surface states can be related by the right-hand-thumb rule; that is, the right-hand-thumb points to the direction of VCVs, the curling fingers point to the circulating direction of the valley surface states (see Fig. 4(a)). Note that the right-hand-thumb rule applies to all sample surfaces (see Fig. 4(a)). Physically, this right-hand-thumb rule is an intrinsic relation between the VCVs and the valley surface states, which are both determined by the bulk Hamiltonian [21].

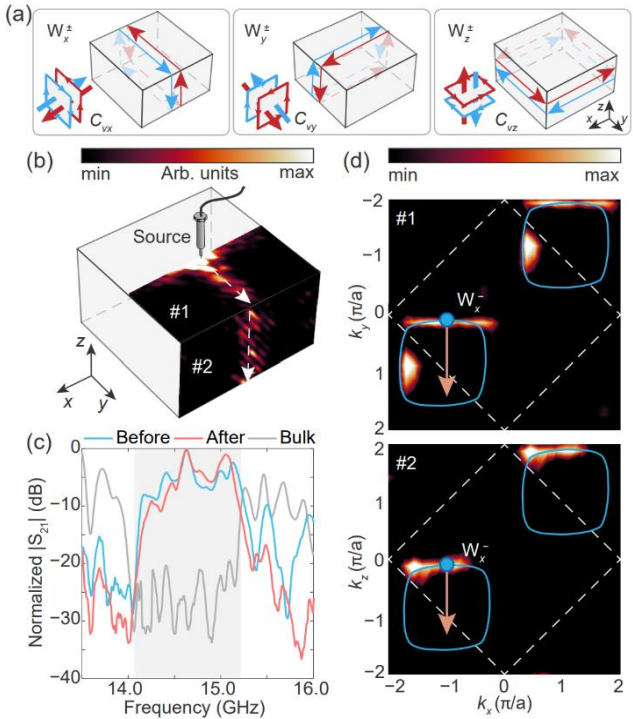

FIG. 4. Right-hand-thumb rule relating the robust chiral valley surface states with the VCVs. (a) Illustration of the direction of VCVs and the propagation loops of valley surface states at three pairs of valleys, respectively. The thin red (blue) arrows denote the $W_\alpha^+$ ($W_\alpha^-$) valley states. The thick arrows in insets denote the direction of VCVs, where the corresponding loops' direction obeys the right-hand-thumb rule. (b) Measured field distribution before (#1) and after (#2) the bend. The dashed lines represent the propagation direction of the surface states. (c) Measured transmission of the bulk states and surface states before and after the bend. The blue and red lines represent surface states before and after the bend, respectively. The gray line represents the bulk states. The shadowed area represents the bulk bandgap. (d) Measured isofrequency contours of the field distribution before (top panel) and after (bottom panel) bend at 14.6 GHz. The carnation arrows represent the direction of the group velocity which is normal to the isofrequency contours. The blue dots represent $W_x^-$ valley. The dashed squares represent the first BZ. The blue lines represent the simulated isofrequency contours.

To experimentally observe the right-hand-thumb rule, we position the source at the center of the (001) surface and measure the field distributions on the different sample surfaces (i.e., (001) and (010) surfaces), as shown in Fig. 4(b). The excited surface states propagate along the $+y$ direction, corresponding to $W_x^-$ valley. Owing to the valley conservation, the surface states smoothly pass through the sharp corner with high transmission (see Fig. 4(c)) and propagate along the $-z$ direction. We also perform Fourier transform on the field distributions on the (001) and (010) surfaces, respectively. As depicted in Fig. 4(d), on both surfaces, only the surface states around the $W_x^-$ valley are excited, experimentally revealing the valley-locking property. Crucially, as the $W_x^-$ valley carries a VCV $(-1/2, 0, 0)$ and the valley surface states circulate clockwise on the $y$-$z$ plane, the direction of the VCV and the circulating direction of the valley surface states obey the right-hand-thumb rule as aforementioned.

## DISCUSSION

In summary, we have developed both theoretically and experimentally the concept of 3D topological valley photonics. From an application perspective, the orbital moments carried by the 3D valley states provide a means to control the angular momentum of light, and the highly directional outcoupling from these valley states can serve as a mechanism for creating narrow-beam antennae. Additionally, the robust 2D canalized valley chiral surface states may inspire innovative 3D topological photonic devices for guiding and trapping light in an unprecedented way. Though demonstrated at lower frequencies, the concept of 3D valley photonics can be readily extended to higher frequencies, including terahertz and optical ranges, where all-dielectric diamond-like 3D photonic crystals with complete photonic bandgaps have already been extensively studied [35].

**ACKNOWLEDGMENTS**

**Funding**: The work was sponsored by the Key Research and Development Program of the Ministry of Science and Technology under Grants 2022YFA1405200 (Y.Y.), No.2022YFA1404704 (H.C.), 2022YFA1404902 (H.C.), and 2022YFA1404900 (Y.Y.), the National Natural Science Foundation of China (NNSFC) under Grants No. 62175215 (Y.Y.), and No.61975176 (H.C.), the Key Research and Development Program of Zhejiang Province under Grant No.2022C01036 (H.C.), the Fundamental Research Funds for the Central Universities (2021FZZX001-19) (Y.Y.), the Excellent Young Scientists Fund Program (Overseas) of China (Y.Y.), and the Start-up Fund and the Direct Grant (Grant No. 4053675) of the Chinese University of Hong Kong (H.X.). **Author contributions**: Y.Y. and H.X. conceived the project. W.L. and Y.Y. carried out the analytical and numerical modeling. W.L., Q.C., and N.H. designed and fabricated the sample. W.L. conducted the measurements with the assistance from F.C. and J.W. W.L. performed data analysis. W.L. wrote the manuscript. W.L., Y.Y., Q.C., H.X., X.L., Y.P., Y.R., and H.C. revised the manuscript. Y.Y., H.X., and H.C. supervised the entire project. All authors discussed the results and commented on the article. **Competing interests**: The authors declare no competing interests. **Data and materials availability**: All data are available from the corresponding authors upon reasonable request.

∗ hansomchen@zju.edu.cn
† haoranxue@cuhk.edu.hk
‡ yangyihao@zju.edu.cn

[1] D. Xiao, W. Yao, and Q. Niu, *Valley-Contrasting Physics in Graphene: Magnetic Moment and Topological Transport*, Phys. Rev. Lett. **99**, 236809 (2007).

[2] J. R. Schaibley, H. Yu, G. Clark, P. Rivera, J. S. Ross, K. L. Seyler, W. Yao, and X. Xu, *Valleytronics in 2D Materials*, Nat Rev Mater **1**, 16055 (2016).

[3] H. Xue, Y. Yang, and B. Zhang, *Topological Valley Photonics: Physics and Device Applications*, Advanced Photonics Research **2**, 2100013 (2021).

[4] J.-W. Liu, F.-L. Shi, X.-T. He, G.-J. Tang, W.-J. Chen, X.-D. Chen, and J.-W. Dong, *Valley Photonic Crystals*, Advances in Physics: X **6**, 1905546 (2021).

[5] X. Wu, Y. Meng, J. Tian, Y. Huang, H. Xiang, D. Han, and W. Wen, *Direct Observation of Valley-Polarized Topological Edge States in Designer Surface Plasmon Crystals*, Nat Commun **8**, 1304 (2017).

[6] J. Noh, S. Huang, K. P. Chen, and M. C. Rechtsman, *Observation of Photonic Topological Valley Hall Edge States*, Phys. Rev. Lett. **120**, 063902 (2018).

[7] F. Gao, H. Xue, Z. Yang, K. Lai, Y. Yu, X. Lin, Y. Chong, G. Shvets, and B. Zhang, *Topologically Protected Refraction of Robust Kink States in Valley Photonic Crystals*, Nat. Phys. **14**, 140 (2018).

[8] J. Ma, X. Xi, and X. Sun, *Topological Photonic Integrated Circuits Based on Valley Kink States*, Laser & Photonics Reviews **13**, 1900087 (2019).

[9] Y. Yang, Y. Yamagami, X. Yu, P. Pitchappa, J. Webber, B. Zhang, M. Fujita, T. Nagatsuma, and R. Singh, *Terahertz Topological Photonics for On-Chip Communication*, Nat. Photonics **14**, 446 (2020).

[10] Y. Zeng et al., *Electrically Pumped Topological Laser with Valley Edge Modes*, Nature **578**, 246 (2020).

[11] D. Smirnova, A. Tripathi, S. Kruk, M.-S. Hwang, H.-R. Kim, H.-G. Park, and Y. Kivshar, *Room-Temperature Lasing from Nanophotonic Topological Cavities*, Light Sci Appl **9**, 127 (2020).

[12] S. Barik, A. Karasahin, S. Mittal, E. Waks, and M. Hafezi, *Chiral Quantum Optics Using a Topological Resonator*, Phys. Rev. B **101**, 205303 (2020).

[13] M. Wang, Q. Ma, S. Liu, R.-Y. Zhang, L. Zhang, M. Ke, Z. Liu, and C. T. Chan, *Observation of Boundary Induced Chiral Anomaly Bulk States and Their Transport Properties*, Nat Commun **13**, 5916

(2022).

[14] W. Li et al., *Topologically Enabled On-Chip THz Taper-Free Waveguides*, Advanced Optical Materials **11**, 2300764 (2023).

[15] T. Ma and G. Shvets, *All-Si Valley-Hall Photonic Topological Insulator*, New J. Phys. **18**, 025012 (2016).

[16] J.-W. Dong, X.-D. Chen, H. Zhu, Y. Wang, and X. Zhang, *Valley Photonic Crystals for Control of Spin and Topology*, Nature Mater **16**, 298 (2017).

[17] X.-D. Chen, F.-L. Zhao, M. Chen, and J.-W. Dong, *Valley-Contrasting Physics in All-Dielectric Photonic Crystals: Orbital Angular Momentum and Topological Propagation*, Phys. Rev. B **96**, 020202 (2017).

[18] X.-T. He, E.-T. Liang, J.-J. Yuan, H.-Y. Qiu, X.-D. Chen, F.-L. Zhao, and J.-W. Dong, *A Silicon-on-Insulator Slab for Topological Valley Transport*, Nat Commun **10**, 872 (2019).

[19] M.-D. Liu, H.-H. Chen, Z. Wang, Y. Zhang, X. Zhou, G.-J. Tang, F. Ma, X.-T. He, X.-D. Chen, and J.-W. Dong, *On-Chip Topological Photonic Crystal Nanobeam Filters*, Nano Lett. **24**, 1635 (2024).

[20] X. Xi, J. Ma, S. Wan, C.-H. Dong, and X. Sun, *Observation of Chiral Edge States in Gapped Nanomechanical Graphene*, Sci. Adv. **7**, eabe1398 (2021).

[21] *Supplementary Material for "Three-Dimensional Topological Valley Photonics"*.

[22] J. Lu, C. Qiu, M. Ke, and Z. Liu, *Valley Vortex States in Sonic Crystals*, Phys. Rev. Lett. **116**, 093901 (2016).

[23] R. Xi, Q. Chen, Q. Yan, L. Zhang, F. Chen, Y. Li, H. Chen, and Y. Yang, *Topological Chiral Edge States in Deep-Subwavelength Valley Photonic Metamaterials*, Laser & Photonics Reviews **16**, 2200194 (2022).

[24] B. Zhou, H.-Z. Lu, R.-L. Chu, S.-Q. Shen, and Q. Niu, *Finite Size Effects on Helical Edge States in a Quantum Spin-Hall System*, Phys. Rev. Lett. **101**, 246807 (2008).

[25] P. A. Belov, C. R. Simovski, and P. Ikonen, *Canalization of Subwavelength Images by Electromagnetic Crystals*, Phys. Rev. B **71**, 193105 (2005).

[26] G. Hu et al., *Topological Polaritons and Photonic Magic Angles in Twisted α-MoO3 Bilayers*, Nature **582**, 209 (2020).

[27] O. Yermakov, V. Lenets, A. Sayanskiy, J. Baena, E. Martini, S. Glybovski, and S. Maci, *Surface Waves on Self-Complementary Metasurfaces: All-Frequency Hyperbolicity, Extreme Canalization, and TE-TM Polarization Degeneracy*, Phys. Rev. X **11**, 031038 (2021).

[28] D. N. Basov, R. D. Averitt, and D. Hsieh, *Towards Properties on Demand in Quantum Materials*, Nature Mater **16**, 1077 (2017).

[29] L. Lu, C. Fang, L. Fu, S. G. Johnson, J. D. Joannopoulos, and M. Soljačić, *Symmetry-Protected Topological Photonic Crystal in Three Dimensions*, Nature Phys **12**, 337 (2016).

[30] G.-G. Liu et al., *Topological Chern Vectors in Three-Dimensional Photonic Crystals*, Nature **609**, 925 (2022).

[31] Y. Yang, Z. Gao, H. Xue, L. Zhang, M. He, Z. Yang, R. Singh, Y. Chong, B. Zhang, and H. Chen, *Realization of a Three-Dimensional Photonic Topological Insulator*, Nature **565**, 622 (2019).

[32] A. Slobozhanyuk, S. H. Mousavi, X. Ni, D. Smirnova, Y. S. Kivshar, and A. B. Khanikaev, *Three-Dimensional All-Dielectric Photonic Topological Insulator*, Nature Photon **11**, 130 (2017).

[33] H. Wang, L. Xu, H. Chen, and J.-H. Jiang, *Three-Dimensional Photonic Dirac Points Stabilized by Point Group Symmetry*, Phys. Rev. B **93**, 235155 (2016).

[34] M. Kim, Z. Jacob, and J. Rho, *Recent Advances in 2D, 3D and Higher-Order Topological Photonics*, Light Sci Appl **9**, 130 (2020).

[35] J. D. Joannopoulos, editor, *Photonic Crystals: Molding the Flow of Light*, 2nd ed (Princeton University Press, Princeton, 2008).